\documentclass[aip, pop, amsmath, amssymb, reprint,]{revtex4-1}

\usepackage{graphicx}
\usepackage{dcolumn}
\usepackage{bm}
\usepackage[utf8]{inputenc}
\usepackage[T1]{fontenc}
\usepackage{mathptmx}

\newcommand{\eg}{e.g., }
\newcommand{\ie}{i.e., }

\begin{document}

\title{Rate equations model for multiple magnetic mirrors in various thermodynamic scenarios}

\author{Tal Miller}
\email{talmiller@gmail.com}
\affiliation{Racah Institute of Physics, The Hebrew University of Jerusalem, Jerusalem, 91904 Israel}
\affiliation{Rafael Plasma Laboratory, Rafael Advanced Defense Systems, POB 2250, Haifa, 3102102 Israel}

\author{Ilan Be'ery}
\email{ilanbeery@gmail.com}
\affiliation{Rafael Plasma Laboratory, Rafael Advanced Defense Systems, POB 2250, Haifa, 3102102 Israel}

\author{Ido Barth}
\email{ido.barth@mail.huji.ac.il}
\affiliation{Racah Institute of Physics, The Hebrew University of Jerusalem, Jerusalem, 91904 Israel}

\date{21 October, 2021}

\begin{abstract}

Axial particle loss is one of the main challenges for fusion aimed, linear magnetic mirror plasma configurations.
One way to mitigate this disadvantage and increase the confinement time is to use a multiple mirrors setup.
The idea is to reduce the outgoing flux by collisions in the outer magnetic cells. 
Here, we develop a semi-kinetic rate equation model for the ions' density dynamics, including scattering within the magnetic cell and the transmission between neighboring cells.
The dominant parameter is the ions' mean free path, which depends on the temperature and density in each cell.
The steady-state flow is studied analytically and numerically for three thermodynamic scenarios: isothermal plasma, adiabatic expansion, and constant diffusion. 
It is found that the confinement time varies about five-fold over the different scenarios, where the adiabatic cooling is the best confining scenario.

\end{abstract}

\maketitle

\section{Introduction}
The mainstream in magnetically confined fusion-aimed systems nowadays relies on closed toroidal geometry. 
Nonetheless, the concept of open, linear magnetic traps might be still attractive for fusion because of its engineering simplicity, continuous operation, the small number of instability modes, and the potential for high efficiency in the use of the magnetic field (high $\beta$, the ratio of plasma pressure to magnetic pressure). 
The two major challenges in mirror machines are flute instability and escaping particles through the loss cones. 
While the flute instability can be controlled with passive,\cite{tajima1991instabilities, beklemishev2010vortex, ryutov2011magneto} RF,\cite{ferron1983rf, seemann2018stabilization, ryutov2011magneto} and active \cite{zhil1975plasma, be2015feedback, ryutov2011magneto} methods, the loss cone flux limits the achievable fusion gain to a non-practical level in simple linear machines. 
In the past, several magnetic configurations have been suggested in order to reduce the axial outward flux including, tandem plugs with thermal barriers,\cite{inutake1985thermal, grubb1984thermal,pratt2006global, tamano1995tandem}, diamagnetic confinement \cite{beklemishev2016diamagnetic, kotelnikov2020structure}, multi-mirrors (MM) systems,\cite{post1967confinement, logan1972multiple, logan1972experimental, mirnov1972gas, makhijani1974plasma, tuszewski1977transient, burdakov2016multiple, budker1971influence, mirnov1996multiple, kotelnikov2007new, mirnov1972gas} moving multiple mirrors,\cite{tuck1968reduction, budker1982gas}, and helical mirror with rotating plasma.\cite{beklemishev2013helicoidal, postupaev2016helical, sudnikov2019first}
Here, we will focus on the MM configuration.

MM systems were studied within different plasma models including, one dimensional diffusion,\cite{makhijani1974plasma, kotelnikov2007new} magneto-hydrodynamics,\cite{makhijani1974plasma} kinetic models,\cite{mirnov1972gas, makhijani1974plasma, matsuda1986relativistic, killeen2012computational, yurov2016nonstationary} and single particle Monte-Carlo method.\cite{logan1972multiple, makhijani1974plasma} 
The main parameter dictating its performance is the collisionality, defined as $\left(\lambda/l\right)^{-1}$, where $\lambda$ is the ions mean free path (MFP) and $l$ is the length of a single mirror cell (more precisely we should use $l_B\sim B/\frac{\partial B}{\partial z}$, the scale over which the magnetic field of the mirror changes, but usually $l_B \approx l$).  
MM systems rely on the MFP being of the order of the cell length, \ie $\lambda/l \approx 1$, for optimal operation, so a diffusion-like dynamics takes place. 
If the MFP is too large, the particles will pass through all the cells uninhibited.
On the other hand, if the MFP is too short, many collisions occur during one bouncing, so the magnetic moment can no longer be regarded as an adiabatic invariant, and the magnetic mirror mechanism ceases working.
The concept of MM has already been verified experimentally both in small and cold systems \cite{logan1972experimental} and more recently in large and hot systems \cite{burdakov2016multiple} but never tested in large and hot, fusion-like, conditions. 

The literature usually assumes that the plasma is isothermal throughout the MM system.\cite{makhijani1974plasma}
However, several experiments exhibit temperature gradients in similar systems.\cite{astrelin1998generation, arzhannikov2003direct, sheehan2014temperature, Wetherton2021} 
This effect can be understood as an adiabatic expansion, much like neutral gas cooling down while expanding through a tube pressurized on one end. 
Therefore, it is interesting to study this cooling effect on the behavior of open magnetic systems such as MM. 
Moreover, since the ions MFP scales as $\lambda\sim T^2/n$, where $T$ is the temperature, and $n$ is the ions density,\cite{goldston1995introduction} the temperature profile significantly affects both the dynamics and the steady-state.
Briefly, in isothermal plasmas, the MFP increases when density decreases, so the plasma becomes less collisional at the ends.
In this case, the trapping of particles in the outer cells is less efficient than in the inner cells.
The opposite happens in the cooling regime, where the temperature decreases with the density, reducing the MFP, so the plasma becomes more collisional at the outer cells, resulting in a more efficient confinement.
This scenario seems to be attractive and is studied in this paper for different cooling models.
For comparison, we also study the isothermal scenario and a third thermodynamic scenario that assumes constant MFP throughout the MM system. 

To this end, we develop in this paper a semi-kinetic model of rate equations for MM systems, as schematically illustrated in Fig. \ref{fig: MM schematic}, and study the steady-state operation for various thermodynamic regimes.
The model includes phase space scattering in each cell and particle exchange between the cells.
The MFP, which determines the scattering rates, depends on both density and temperature that are self-consistently calculated in our model.
The resulting steady-state axial density profiles and the outgoing flux, which is inversely proportional to the confinement time, are calculated for different thermodynamic regimes.
We employ our rate equation model to study all three thermodynamic scenarios, including three different models for the adiabatic cooling (see below).
The comparison between the three scenarios reveals that the steady-state density profiles are significantly different, and the confinement time varies by up to an order of magnitude, where the best scenario is the adiabatic cooling, which is studied here for the first time.
Two regimes of operation are studied, $\lambda/l\approx1$ and $\lambda/l\gg 1$, where we refer to the first as the optimal regime and the latter as the sub-optimal regime.

The numerical results are compared with a theory based on the diffusion equation for all thermodynamical scenarios and exhibit a very good agreement.
Our results also comply with the predicted $1/N$ scaling of the outgoing flux\cite{logan1972multiple, makhijani1974plasma} ($N$ being the number of MM cells) 
as well as with the $1/R_m$ scaling\cite{kotelnikov2007new}, where $R_m$ is the mirror ratio.

The structure of the paper is as follows.
Section \ref{model} introduces the rate equations model and discusses the possible thermodynamic regimes. 
Section \ref{results} presents the system's parameters and the numerical results for the different thermodynamic scenarios. 
The results are compared with the diffusion-based theory that is developed in Sec. \ref{theory}.
The conclusions are summarized in Section \ref{conclusions}.

\section{The Rate Equations Model}
\label{model}
\subsection{Assumptions}

\begin{figure}[tb]
    \centering
    \includegraphics[scale=0.38]{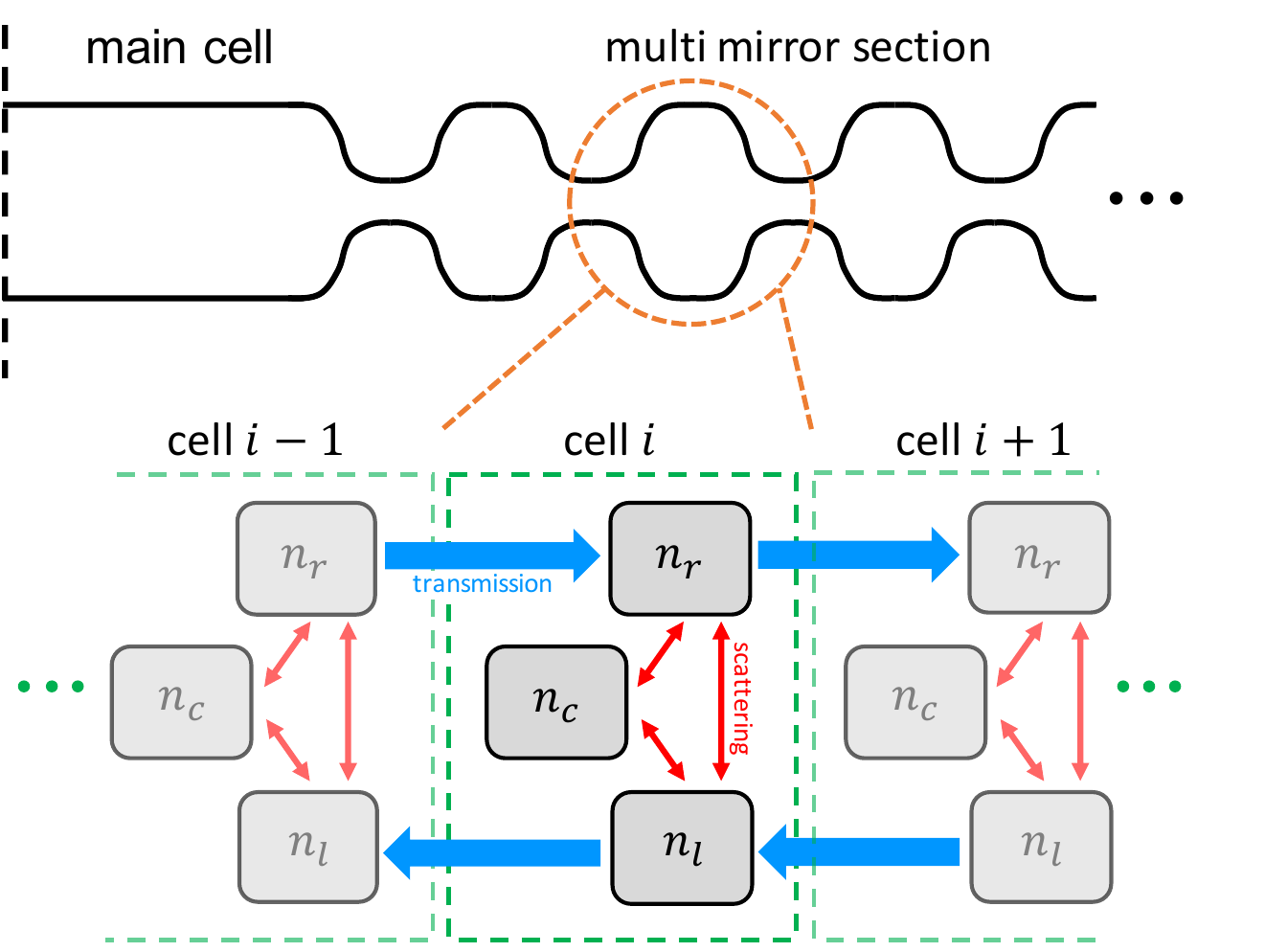}
    \caption{Schematic drawing of the MM system and the particle populations described by the rate equations model. The arrows represent particle transfer between different populations in and between MM cells.}
    \label{fig: MM schematic}
\end{figure}

Our main goal is to study the effect of adding MM sections to the central mirror cell on the plasma confinement efficiency in a linear magnetic mirror configuration as schematically illustrated in Fig \ref{fig: MM schematic}. 
To this end, we develop a simplified model of rate equations for the densities in each MM cell.
By solving the (nonlinear) rate equations, we can find the steady-state solution for the densities profiles and calculate the outgoing flux of escaping particles.

For each magnetic mirror cell, $i$, we identify three populations in phase space, (a) trapped particles due to the mirror effect, (b) right going (through the loss cone) particles from cell $i$ to cell $i+1$, and (c) left going particles from cell $i$ to cell $i-1$.  
The densities of these populations are denoted by captured $n_c$, right going $n_{r}$ and left going $n_{l}$, respectively, and the total ion density is $n=n_c+n_{r}+n_{l}$.
The system comprises a central cell for fusion (the main cell) and two, right and left, MM sections for reducing the outgoing particle flux.
We choose to look at the right MM section.
Thus, $n_r$ is the escaping population, while $n_l$ is the population that returns toward the central cell as illustrated in Fig \ref{fig: MM schematic}.  
Mathematically speaking, the model comprises three rate equations for the three populations in each cell, so overall, we have $3N$ ordinary differential equations.
For completeness, boundary conditions that are required for the inward fluxes at the two ends are defined in Sec. \ref{BC}. The main assumptions in the model are as follows:

First, we consider only the dynamics along the mirror axis and neglect all radial effects such as radial diffusion and instabilities.
Therefore, our model includes the effects of particle transfer between cells and scattering within each cell that mixes between the trapped and the untrapped populations. 
Also, our model describes the averaged dynamics in each mirror cell and disregards fluctuations.

Second, the discrete description of the system fails when each MM cell is not well defined, and the dynamics in the transmission region between neighboring cells become significant. 
Therefore, we assume $R_m\gg1$, where $R_m=B_{max}/B_{min}$ is the mirror ratio.

Third, since the mobility of the ions is much smaller than that of the electrons along the mirror axis (unlike the radial axis where it is the opposite due to the ions' larger cyclotron radius), the ions dominate the axial transport properties of the plasma, where the electrons follow the ions to keep quasi-neutrality.
Therefore, we consider in the model the ion dynamics only and disregard the electrons.
However, the electrons may affect the thermodynamics properties of the systems by accelerating the thermalization between the systems' cells, as will be addressed below.   
Another effect of the electrons is the ambipolar electric field that increases the diffusion coefficient by a factor of about $f_{ap}=1+T_e/T_i$, for systems without temperature gradients, where $T_i$ and $T_e$ are the ion and electron temperatures, respectively. \cite{makhijani1974plasma} 
In other words, the ions dominate the plasma dynamics, but the effective transport rates are affected by the background electrons. 
In our simulations, we use $f_{ap}=2$ (corresponding to equal ion and electron temperatures) even for the non-isothermal scenarios. 
However, changing the factor does not qualitatively change the results (dividing it by 2 reduces the outward particle flux, defined in section \ref{steady state}, by less than a factor of 2).
Thus, we drop below the ion/electron subscripts and refer strictly to the ions unless stated otherwise.

Besides, for a symmetric mirror, the loss cones (left or right) are defined in velocity space as $\left(v_{\perp}/v\right)^{2}<R_{m}^{-1}$, where, 
$v$ is the total velocity, $v_{\perp}$ is the perpendicular velocity.\cite{stacey1981fusion} 
The opening angle, $\theta$, of the loss cones with respect to the mirror axis satisfies $\sin\theta=v_{\perp}/v=R_{m}^{-1/2}$. 
The loss cone solid angle is then $\Omega=4\pi\sin^{2}\left(\theta/2\right) $, so the fraction of each (left or right) loss cone out of the total velocity space is $\alpha=\Omega/4\pi$. 
For small angles (or large $R_m$) it can be approximated as $\alpha \approx \left( 4 R_{m} \right) ^{-1}$. 
Therefore, the parameter $\alpha$ determines the probability of a scattered particle ending up inside one of the loss cones ($\alpha$) or outside the loss cones ($1-2\alpha$).
It is noted that a more precise model may consider a smaller available phase space region for the scattered particles, as was assumed by Skovorodin and Beklemishev. \cite{skovorodin2012plasma}
We have checked the influence of such an assumption and found (not presented) that the relative difference in the confinement time between the different thermodynamics regimes (Sec. \ref{scenarios}) is less than $10\%$.
Since our main result regarding the confinement improvement in the adiabatic cooling scenario (see Sec. \ref{examples}) is insensitive to this assumption refinement, we do not include it our rate equations model, which is simplified but sufficient to capture the main effect.

Finally, Kotelnikov\cite{kotelnikov2007new} analyzed various regimes of the MM confinement time in the four-dimensional parameter space, $\{l,L,R_m,\lambda\}$, for isothermal plasmas, where $L=Nl$ is the system length. 
However, most of these regimes considered weak $(R_m-1 \ll 1)$ or mild $(R_m-1 \approx 1)$ ripples inconsistently with the assumptions underlying our rate equations model mainly due to its discrete nature. 
Nevertheless, the "strong ripples" regimes, $R_m-1 \gg 1$, comply with the rate equations assumptions.
Thus, we compare in Sec. \ref{scaling with Rm} the predicted linear scaling of plasma lifetime with the mirror ratio with our rate equations results finding an excellent agreement.

\subsection{Rate Equations}
\label{rate eqs}

We write rate equations for the densities of the three populations of each cell.
The processes included in the model are the Coulomb scattering within each cell and transmission through the loss-cones. 
Each process has a characteristic time scale that is inversely proportional to the relevant population's density change rate.

The first process in the model is the ion-ion Coulomb scattering rate, given by 
\begin{eqnarray}
    \nu_{s} = \frac{1}{12\pi^{3/2}}\frac{Z^{4}e^{4}n\ln\Lambda_{ii}}{\epsilon_{0}^{2}m^{1/2}\left(k_{B}T\right)^{3/2}},
\end{eqnarray}
where $k_{B}$ is the Boltzmann constant, $\epsilon_{0}$ the vacuum permittivity, $m$ the ion mass, $e$ the electron charge, and $Z$ the ion charge number. \cite{fundamenski2007comparison}
$\ln\Lambda_{ii}$ is the ion-ion Coulomb logarithm given by 
\begin{eqnarray}
    \ln\Lambda_{ii}&=&23-\ln\left(Z^{3}\sqrt{2n}\,T^{-3/2}\right),
\end{eqnarray}
where the ions density $n$ is given in units of $\mathrm{cm}^{-3}$ and the ions temperature $T$ is given in units of $\mathrm{eV}$. \cite{richardson2019nrl}

During the scattering time, $\tau_{s}=1/\nu_{s}$, the velocities of the particles get randomized, so we assume isotropic distribution of the scattered particles regardless of the average velocity before scattering.
Therefore, the probability of one population, say $n_1$, to scatter into another population, say $n_2$, is proportional to the fraction of the solid angle of population $n_2$ in velocity space. 
These fractions were defined above as $\alpha$ for each of the transmitting populations, $n_r$ and $n_l$, and $1-2\alpha$ for the captured population, $n_c$. 
For example, the term in the rate equations associated with $n_{l}$ scattering to $n_{c}$ is $\alpha \nu_{s} n_{l}$.
In order to preserve the total number of particles, this term appears with opposite signs in the equations for $\dot{n}_{c}$ and $\dot{n}_{l}$.

The second process is the transmission of particles between neighboring cells through the loss cone.
We approximate the transmission rate $\nu_{t}$ as the ratio between the ion thermal velocity, $v_{th}=\sqrt{3 k_{B}T/m}$ and the cell length $l$, multiplied by the ambipolar factor $f_{ap}$, \ie the transmission rate of cell $i$  is $\nu_t^i = v_{th}^i f_{ap} /l $. 
Each cell transmits particles to and receives particles from its neighboring cells at rate $\nu_{t}$.
Therefore, for example, the term in the rate equations associated with particles transmitted from cell $i$ to cell $i+1$ (via the right loss cone) is $\nu_{t}^i n_{r}^i$.

To summarize, the $3N$ rate equations for the three populations of the $N$ cells in the right half of the MM system, as illustrated in Fig. \ref{fig: MM schematic}, are
\begin{eqnarray}
\label{Eq: dn_c_dt} \dot n_{c}^i &=&  \nu_{s}^i \left[(1-2\alpha) (n_{l}^i + n_{r}^i) - 2\alpha n_{c}^i\right]  \\
\label{Eq: dn_tL_dt} \dot n_{l}^i &=&  \nu_{s}^i\left[\alpha (n_{r}^i+n_{c}^i) - (1-\alpha ) n_{l}^i\right] - \nu_{t}^i n_{l}^i + \nu_{t}^{i+1} n_{l}^{i+1} \; \\ 
\label{Eq: dn_tR_dt} \dot n_{r}^i &=& \nu_{s}^i \left[\alpha (n_{l}^i +n_{c}^i) - (1-\alpha) n_{r}^i\right]- \nu_{t}^i n_{r}^i +  \nu_{t}^{i-1} n_{r}^{i-1} \;
\end{eqnarray}
Cell $1$ is the closest to the central (fusion) cell, and cell $N$ is the outer cell (last to the right) of the system.
The steady-state solution yields the steady-state flux of escaping particles and thus determines the lifetime of the system.
Since the rate coefficients $\nu_s$ and $\nu_r$ depend on each cell's (ion) temperature, the overall behavior of the systems depends on the thermodynamical scenario of the MM system, as will be discussed in Sec. \ref{scenarios}.

\subsection{Boundary Conditions}
\label{BC}

To close the set of rate equations for the MM system, we must include boundary conditions.
Practically, the quantities to be defined are the incoming flux to cell $1$, \ie $\nu_t^0$, and the outgoing flux from cell $N$, $\nu_t^{N+1}$.

For the left boundary condition, we note that $\nu_t^0 n_r^0$ is the source flux from the central (fusion) cell to the first MM cell.
One option is to consider this flux term as a constant that characterizes the loss rate from the central cell.
However, if this quantity were known, we would not need to develop a model to calculate it.
Alternatively, since we will be interested in the steady-state, we consider a condition of constant total density in the first (left) cell, $n_c^1+n_{l}^1+n_{r}^1=n_0$. 
This condition simulates a steady-state, where the total density in the first cell is constant and equals that of the central cell. 
In this state, the injection of new particles into the central fusion cell compensates the constant outgoing particle flow.
Practically, in each time step, we first advance $n_c^1$ and $n_{l}^1$ populations according to the rate equations, and then set $n_{r}^1=n_0-n_{l}^1-n_c^1$.

As to the right boundary condition, we consider a "free flow" condition, which we implement by simply setting $\nu_t^{N+1}n_l^{N+1}=0$.
Now, we have a closed set of rate equations and what is left is to add a model for the temperature profile in the system.

\subsection{Thermodynamic Scenarios}
\label{scenarios}

The main idea behind the MM is to scatter particles out of the loss cone.
Therefore the most important parameter is the system collisionality, $l/\lambda$, where $l$ is the mirror cell length and $\lambda=v_{th}/\nu_{s}$ is the MFP. \cite{post1967confinement, logan1972multiple, logan1972experimental, mirnov1972gas, makhijani1974plasma, tuszewski1977transient, burdakov2016multiple, budker1971influence, mirnov1996multiple, kotelnikov2007new, mirnov1972gas}
Since $v_{th}\propto\sqrt{T}$ and $\nu_s\propto n/T^{3/2}$ (see section \ref{rate eqs}), the MFP scales as $\lambda\propto T^{2}/n$. \cite{goldston1995introduction} 
If the system is isothermal, \ie the temperature is the same in all cells, the MFP increases as the plasma expands, and the density drops towards the outer MM cells.
As a result, the collisionality decreases and drives the system away from the optimal regime.
Indeed, in the literature, MM systems are commonly considered isothermal. \cite{makhijani1974plasma}
However, several experimental works have shown that temperature gradients are possible in expanding magnetized plasmas. \cite{astrelin1998generation, arzhannikov2003direct, sheehan2014temperature} 
These experimental results imply that for certain systems and specific thermodynamical conditions, the temperature gradient might compensate the density gradient such that the MFP would stay constant or even decrease along with the system.
In such scenarios, we expect to see a significant improvement in the overall MM confinement.
However, the exact temperature profile depends on the specific thermodynamical properties of a given system, which is beyond our simplified model.
Therefore, we use our model to study how different thermodynamic scenarios affect the density profiles and the confinement time, leaving the question of determining the correct scenario for a future study.

The simplest cooling model is adiabatic cooling.
Although this scenario neglects the contribution of the electrons to the system thermalization by transporting energy between the system's cells, we will consider it a limiting regime for decreasing MFP.
For ideal gases, the adiabatic cooling law is $T \propto n^{\gamma-1}$ where $\gamma=\left(d+2\right)/d$ and $d$ is number of degrees of freedom. \cite{be2018plasma} 
Plugging this relation in the MFP gives $\lambda \propto n^{2\gamma-3}$ in the cooling scenario. 
Hence, the MFP will decrease on expansion for systems with $d<4$ while an isothermal plasma is associated with the limit $d\rightarrow\infty$. 
Effectively, the longitudinal expansion of plasmas in linear machines, such as the MM systems, is one-dimensional. \cite{stacey1981fusion,bellan2008fundamentals} 
Therefore it is reasonable to consider $d=1$ in our model in the adiabatic regime. 
However, to examine the effect of different effective expansion regimes, we also study the cases with $d=2,3$. 

In addition to the isothermal and adiabatic cooling scenarios, we will also study a thermodynamic scenario with a constant MFP.
This regime corresponds to a one-dimensional diffusion with a constant diffusion coefficient. 
The MFP scaling with density in this model falls between the isothermal and cooling scenarios. 
Therefore we expect it to be between these models also in terms of the confinement improvement.
Besides, although it is unlikely that the collisionality would not change while both density and temperature vary, there may be a way to manipulate the MFP (\eg by external fields) such that it would stay constant throughout the system.
Interestingly, the GOL-3 experiment\cite{arzhannikov2006experimental}, an MM trap with an intense electron beam for plasma heating, exhibited ion bounce instabilities that drove the system into the optimal ratio $\lambda/l \approx 1$ for all MM cells.\cite{beklemishev2007bounce, skovorodin2013flow}
Therefore, this system might correspond to the constant MFP scenario in our rate equations model.

Finally, we note that although we have covered a broad range of possible parameters and thermodynamic scenarios, the rate equations model can be generalized, in a future study, to other systems or modes of operation by modifying the scattering and transmitting coefficients.

\subsection{Steady State}
\label{steady state}

Although the system dynamics can be studied within our rate equations model, we choose to focus on the steady-state and leave the dynamics for a future study.
The steady-state solution obeys $\dot{n}=0$ for all populations in all cells. 
At steady-state, the total (ion) flux between two neighboring cells (\eg $i\rightarrow i+1$),
\begin{eqnarray}
    \phi_{i,i+1}=S\cdot \left( v_{th}^i n_{r}^i -v_{th}^{i+1} n_{l}^{i+1} \right).
\end{eqnarray}
Here, $S$ is the mirror cross-section area, which we assume to be the same throughout the MM section. 
The total particle flux is twice the ion flux since we assume the electrons follow the ions. 
In the steady-state, the (ion) flux should be constant and equal for all $i=1,\dots N-1$, \ie \, $\phi_{i,i+1}=\text{const}=\phi_{ss}$.
Notably, the confinement time of the system is inversely proportional to $\phi_{ss}$.
For constant coefficients, the steady-state can be found analytically by a simple matrix diagonalizing.
However, in general, both $\nu_s$ and $\nu_t$ depend on temperature and density, so we get nonlinear rate equations to be solved numerically.
To this end, we choose to use the relaxation method, in which we start with an initial guess for the populations profiles and evolve the rate equations toward the steady-state.
Practically, we stop the simulations when the standard deviation of the fluxes, $\{\phi_{i,i+1}\}_{i=1}^{N-1}$,  was less than $5\%$ of the mean flux.
The obtained profile is defined to be the (approximated) steady-state.

\section{Simulations}
\label{results}

\subsection{System Parameters}
\label{parameters}

Common plasma parameters considered for a D-T fusion machine are $n=10^{21}\,\mathrm{m^{-3}}$ and $k_{\rm B} T=10\,\mathrm{keV}$. 
Such plasmas would require a magnetic field of at least $B=2\mathrm{T}$ to satisfy the confinement condition, $\beta\leq1$, where $\beta$ is the ratio between the plasma pressure,  $P = 2 n k_B T$ (where $n$ stands for the ion density, so the total density is twice that), and the magnetic pressure, $P_M=B^2/2\mu_0$. 
The ion MFP, in this case, is $\lambda \approx 1800\mathrm{m}$, which is nowhere near practical for MM systems if we want to be in the efficient regime of $l\approx \lambda$.
In order to bring the MM system, which is based on collisions, to a more practical set of parameters, one would like to reduce the MFP ($\sim T^2/n$) without changing the fusion power, which scales as $W_{\mathrm{fus}}\sim n^{2}\left<\sigma v\right>\sim n^{2}\xi^{2}\exp\left(-\xi\right)$, where,
$\left<\sigma v\right>$ is the Maxwell-Boltzmann averaged fusion reactivity and $\xi\propto T^{-1/3}$ is a dimensionless parameter. \cite{atzeni2004physics}
Practically, this can be done by increasing the density and reducing the temperature.

In the current work, we study two sets of parameters.
The first set considers $n=2\cdot  10^{22}\, \mathrm{m^{-3}}$ and $k_{\rm B} T=3\mathrm{keV}$, resulting in a MFP of $\lambda=10\mathrm{m}$.
We pick a manageable length of $l=10\text{m}$ for each MM cell, so the plasma in the first MM cell (before the effects of expansion and cooling) is in the optimal (efficient) regime $\lambda/l \approx 1$.
The magnetic field required for this example will be at least $B=10\mathrm{T}$ (for $\beta=0.5$).
For the mirror ratio, we usually considered $R_m=10$ while the effect of varying $R_m$ is studied in Sec. \ref{scaling with Rm}.
We note that the peak magnetic field in the mirror throats in this example is $100\mathrm{T}$, exceeding the current engineering capabilities, but it is used for simplicity while the primary results also hold for smaller  $R_m$ (see Sec. \ref{scaling with Rm}).

The second set of parameters is for the sub-optimal regime. 
The parameters were as in the optimal regime but,the ion density is reduced by a factor of 20 to $n=10^{21}\text{m}^{-3}$.
In this case, besides the less efficient fusion rates, the plasma becomes much less collisional with $\lambda/l=20$, so we expect a more modest confinement improvement in the MM system. 
We note that the parameters range chosen for the GOL-NB machine,\cite{postupaev2016gol, postupaev2019results}   $1\le\lambda/l<N$, covers our both optimal and sub-optimal regimes.
Therefore, the confinement efficiency and the effect of different thermodynamical scenarios for this system can, in principle, be estimated from our rate equation model.

Before presenting the simulation results, let us estimate the expected flux of escaping particles in a simple mirror system (without MM sections) with the above parameters and compare it to the maximal flux dictated by the Lawson criterion.
A naive estimate for the outgoing (ion) flux from the central cell is given by $\phi_{naive}=n v_{th}S$, where $S$ is the cross-section of the plasma at the minimum magnetic field between mirrors.
This estimation is valid in the regime where the MFP is close or small compared to the central cell length, precisely the regime relevant for employing MM, while when the MFP is much larger, the naive flux would be dictated by the Coulomb scattering rate in the central cell.
For plasma diameter of one meter, the fluxes for the optimal parameters regime reads $\phi_{naive}=9.2\cdot10^{27}\text{s}^{-1}$ and for the sub-optimal regime $\phi_{naive}=4.6\cdot10^{26}\text{s}^{-1}$.
These estimations do not include the (single) mirror effect, which reduces the flux by an additional factor of the order of $R_m$,\cite{makhijani1974plasma,Anikeev1999MainRO} which is $\sim 10$.

On the other hand, the Lawson criterion\cite{lawson1957some} for the minimal confinement time is
\begin{eqnarray}
    \tau_{\rm Lawson}=\frac{12k_{B}T}{n\left\langle \sigma v\right\rangle E_{ch}},
\end{eqnarray}
where, $E_{ch} $ is the charged fusion products energy (\eg\,$3.5$  MeV for D-T reactions) and $\left\langle \sigma v\right\rangle$ is the Maxwell-Boltzmann averaged fusion reactivity. 
Since the central cell loses particles from both ends, the maximal (ion) flux out of each end allowed by the Lawson criterion reads
\begin{eqnarray}
    \phi_{Lawson}=\frac{nV}{2\tau_{\rm Lawson}}=\frac{n^{2}\left\langle \sigma v\right\rangle E_{ch}V}{24k_{B}T}
\end{eqnarray}
where, $V$ is the volume of the fusion cell.
As a limit for a practical fusion cell, we consider cell length of $100 m$ and a diameter of one meter, so the plasma volume in the central cell is $V\approx 80 m^3$.
Under these assumptions, the Lawson flux becomes
$\phi_{Lawson}\approx 3\cdot10^{23}\text{s}^{-1}$ for the optimal regime and $\phi_{Lawson}\approx 8\cdot10^{20}\text{s}^{-1}$ for the sub-optimal regime.
Thus, after adding the mirror effect of the central fusion cell (but before taking into account the MM effect), one finds that the estimations for the outgoing flux are around 3 and 5 orders of magnitude higher than the maximal Lawson flux for the optimal and sub-optimal regimes, respectively. 

It is important to note that for mirror machines with much higher ion temperatures (in the range $100-250$KeV) and lower densities (to satisfy $\beta<1$), the Lawson criterion can be achievable\cite{fowler1966fusion, futch1972multi}.
However, the MM approach is not applicable for such high temperatures because the MPF of the ions is of the order of tens of kilometers, while practical MM operation requires MFP of tens of meters at the most.
Therefore, such systems are out of the scope of the current study.

In the following sections, we use the rate equations to calculate the axial density profiles and the corresponding outgoing flux in MM systems focusing on how different thermodynamic scenarios affect the containment time in MM systems.
We will see MM can improve the confinement time by an additional one or two orders of magnitudes where the best scenario is the adiabatic cooling.

Finally, let us estimate the limitations imposed by radial losses, which, as mentioned above, are not included in our rate equations model.
To this end, we compare the axial and the radial diffusion in the MM sections.
Different models for the radial diffusion vary between the so-called classical diffusion,\cite{chen1984introduction} $D_{\mathrm{class}}= r_{\rm gyro}^2 \nu_s$, and Bohm diffusion,\cite{bohm1949characteristics} $D_{\mathrm{Bohm}}=k_B T/16 e B$, 
where, $r_{\rm gyro}$ is the gyroradius.
However, determining the accurate model for a given system in a given thermodynamical conditions remains an open question.
As a rough estimation for our considered parameters, we found that for an MM section with $10$ cells of $10$ meters each in the optimal regime $\lambda/l\approx 1$, the radial diffusion in the MM section is of the order of one percent of the axial  flux for classical diffusion and an order of magnitude larger for Bohm diffusion.
When the radial diffusion becomes of the order of the longitudinal diffusion, one cannot neglect its effect, where for the above parameters, this happens for $N\sim 100$.
Of course, these estimations depend, among other effects, on the exact temperature and density profiles in the system that are determined by the thermodynamical scenario.
For example, in the adiabatic cooling scenario, the classical diffusion might become dominant for longer systems because both temperature and MFP decrease with the cell number (see Fig \ref{fig: MFP profiles}), and the system becomes much more collisional. 
In contrast, the Bohm diffusion decreases in this scenario due to the cooling.  
Therefore, more theoretical and experimental research is needed to estimate the radial diffusion effect for a given system quantitatively.

\begin{figure}[tb]
    \includegraphics[clip, trim=.7cm .5cm .5cm 0.5cm, 
    width=.49\linewidth] {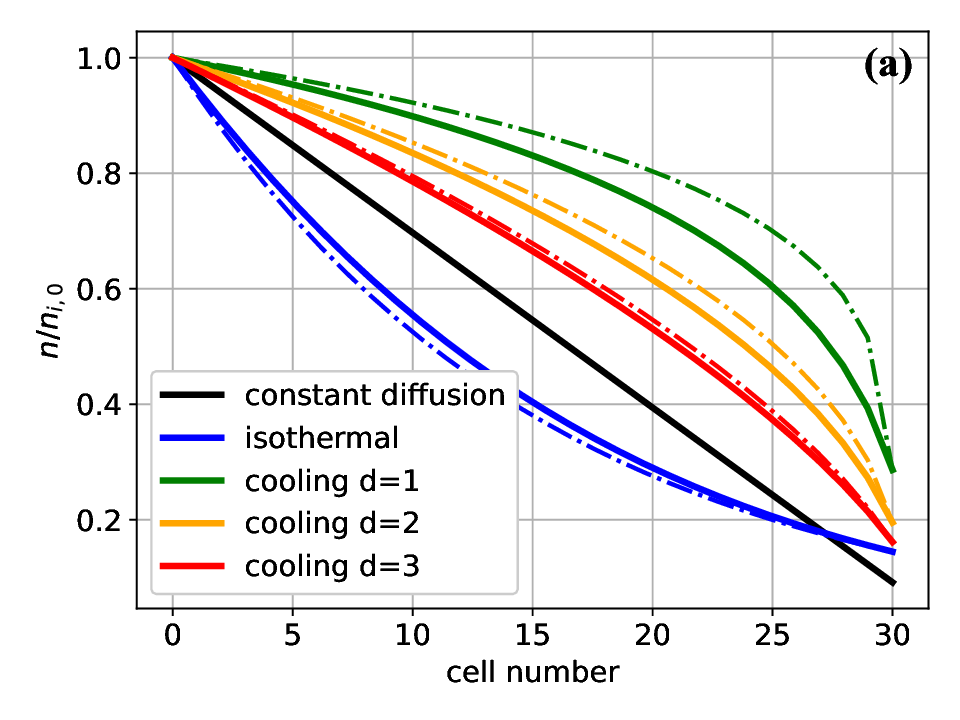}
    \includegraphics[clip, trim=.7cm .5cm .5cm 0.5cm, 
    width=0.49\linewidth] {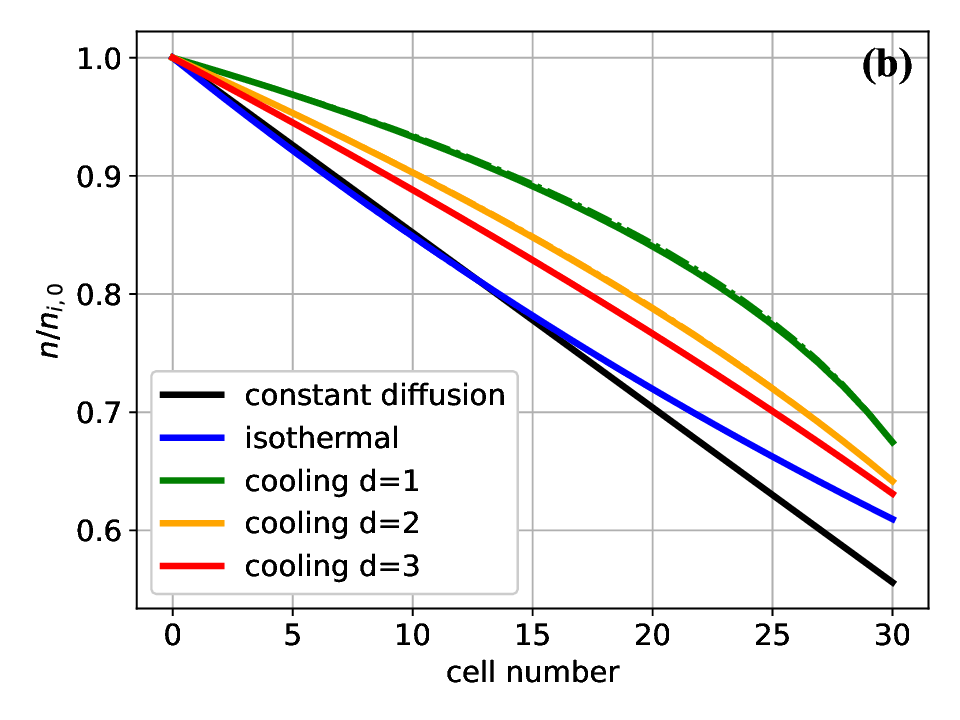}
    \caption{
    Steady-state density profiles (solid lines) for the different thermodynamic scenarios and the two regimes: (a) the optimal regime, $\lambda/l=1$ in the first cell and (b) the sub-optimal regime, $\lambda/l=20$ in the first cell are compared to the theoretical profiles of Sec. \ref{theory} (dash-dotted lines of the same color). 
    The parameters here were $l=10\text{m}$, $k_{\rm B} T=3 \text{keV}$,  $B=10\text{T}$, $R_m=10$, and $n=2\cdot 10^{22}\text{m}^{-3}$ for the optimal regime, and $n= 10^{21}\text{m}^{-3}$ for the sub-optimal regime.}
    \label{fig: density profiles}
\end{figure}

\subsection{Steady State Results \label{examples}}

\begin{figure}[tb]
    \includegraphics[clip, trim=0.7cm 0.5cm 0.5cm 0.2cm,
    width=0.495\linewidth]{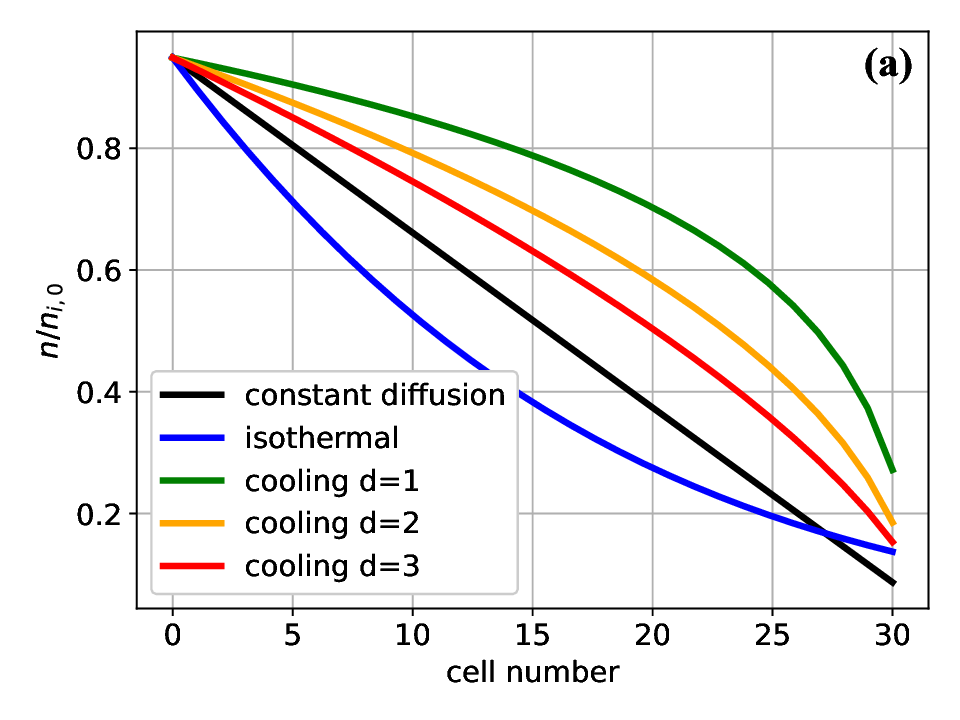}
    \includegraphics[clip, trim=0.7cm 0.5cm 0.5cm 0.2cm, width=0.495\linewidth]{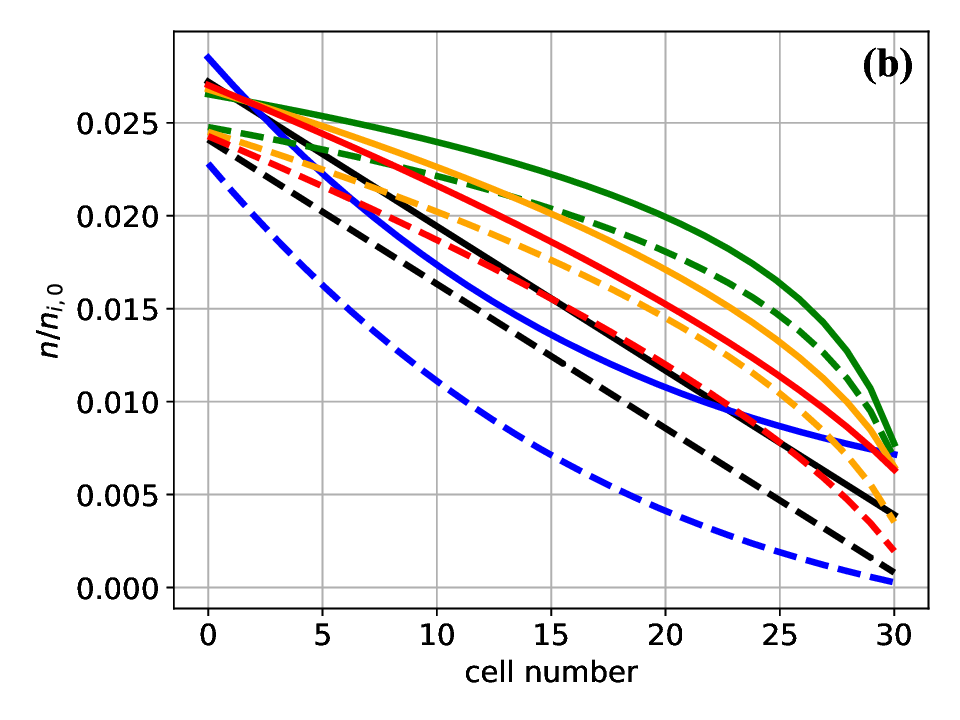}
    \caption{
    Steady-state density profiles of the captured population (a) and the transmitting (right-going in solid lines and left-going in dashed lines) populations (b) for the different thermodynamic scenarios (different colors), where the first cell is in the optimal regime, $\lambda/l=1$ in the first cell. 
    The parameters here were as in Fig. \ref{fig: density profiles}a.}
    \label{fig: population density profiles}
\end{figure}

First, let us consider a MM system with $N=30$ cells and physical parameters described in the previous section.
We have solved the rate equations for both the optimal $\lambda/l=1$ (left) and the sub-optimal $\lambda/l=20$ (right) regimes.
The resulted total steady-state axial density profiles, $n$, are presented in Fig. \ref{fig: density profiles}. 
In each regime, we compare the different thermodynamical scenarios.
It is shown that the constant diffusion scenario (black line) has a linearly declining profile while the other scenarios have curved density profiles.
The diffusion theory that is developed in Sec. \ref{theory} (dashed-dotted lines) remarkably agrees with the rate equations model (solid lines) for $\lambda/l=20$ (right panel), where the differences can hardly be distinguished for all thermodynamical scenarios.
However, in the efficient regime, $\lambda/l=1$, the assumption for continuous diffusion, $\lambda\gg l$, breaks, and the theoretical and the numerical profiles differ mainly for adiabatic scenarios with $d=1,2$. 

In Fig. \ref{fig: population density profiles}, we decompose the total density in each thermodynamical scenarios into its three sub-populations for the optimal regime, $\lambda/l=1$.
In agreement with the considered mirror ratio, about $97\%$ of the particles in each cell are trapped for all thermodynamic scenarios. 
The population difference $n_{r}-n_{l}$ is proportional to the steady-state flux and, therefore, does not depend on the cell number for all scenarios.
We note that in the constant diffusion scenario, the profiles of all three sub-populations decline linearly, while in the other scenarios, the profiles are curved.
This is because the scattering and the transmission rates in these scenarios depend on the density, where the density profile is convex in the isothermal scenario and concave in the three cooling scenarios. 
It is noted that some past experimental works seem to be inconsistent with a concave density profile \cite{logan1972experimental, danilov1975plasma} but more detailed experiments are required to study this problem in different conditions.

\begin{figure}[tb]
    \includegraphics[clip, trim=.5cm 0.5cm .5cm 0.2cm,
    width=0.495\linewidth]{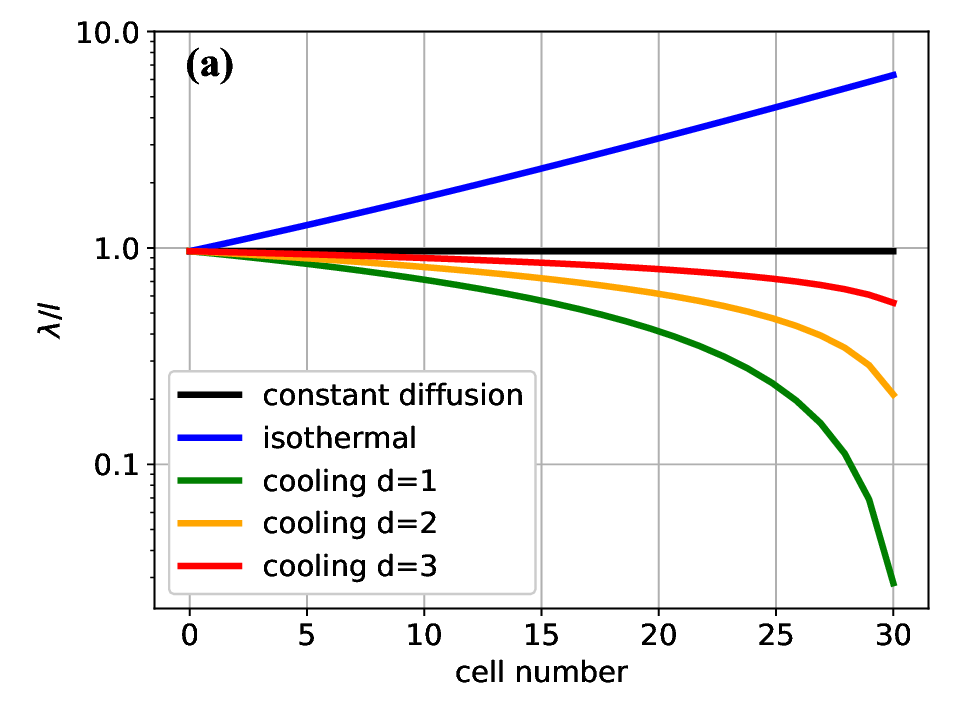}
    \includegraphics[clip, trim=.5cm 0.5cm 0.5cm 0.2cm,
    width=0.495\linewidth]{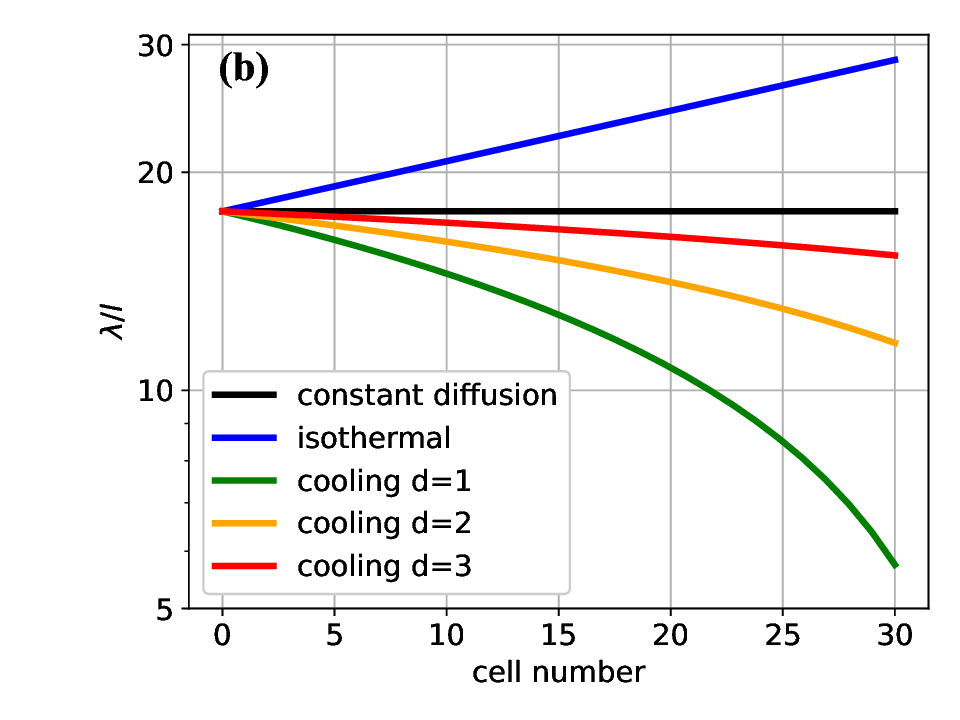}
    \caption{Steady-state mean free path profiles (normalized by the cell length) for a MM system with $N=30$ cells at the different thermodynamic scenarios in the optimal regime, $\lambda/l=1$ in the first cell, (a) and in the sub-optimal regime, $\lambda/l=20$ in the first cell, (b). The parameters in subplots (a) and (b) were as in the subplots of Fig. \ref{fig: density profiles}, respectively.}
    \label{fig: MFP profiles}
\end{figure}

Fig. \ref{fig: MFP profiles} presents the MFP profiles for both optimal (left) and sub-optimal (right) regimes.
As expected, the MFP increases with the cell number for the isothermal scenario and decreases for the cooling scenarios, where the lower the dimension in the cooling scenarios, the lower the MFP gets. 
It is notable that when $\lambda \ll l$ (as happen for large values of $N$ for the adiabatic $d=1$ cooling scenario), the system becomes collisional so the validation of our simplified model for the mirror is questionable, and the results in this scenario must be taken with caution.

\subsection{Scaling With System Size}
\label{scaling with N}

\begin{figure}[b]
    \includegraphics[clip, trim=.5cm 0.5cm .5cm 0.5cm,
    width=0.495\linewidth]{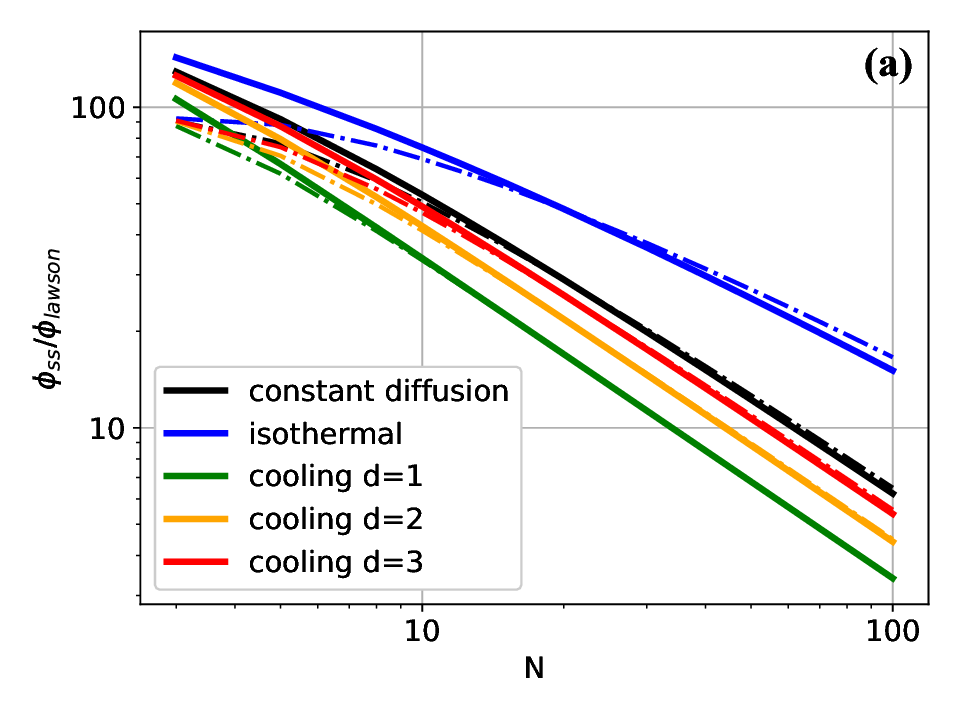}
    \includegraphics[clip, trim=.5cm 0.5cm 0.5cm 0.5cm,
    width=0.495\linewidth]{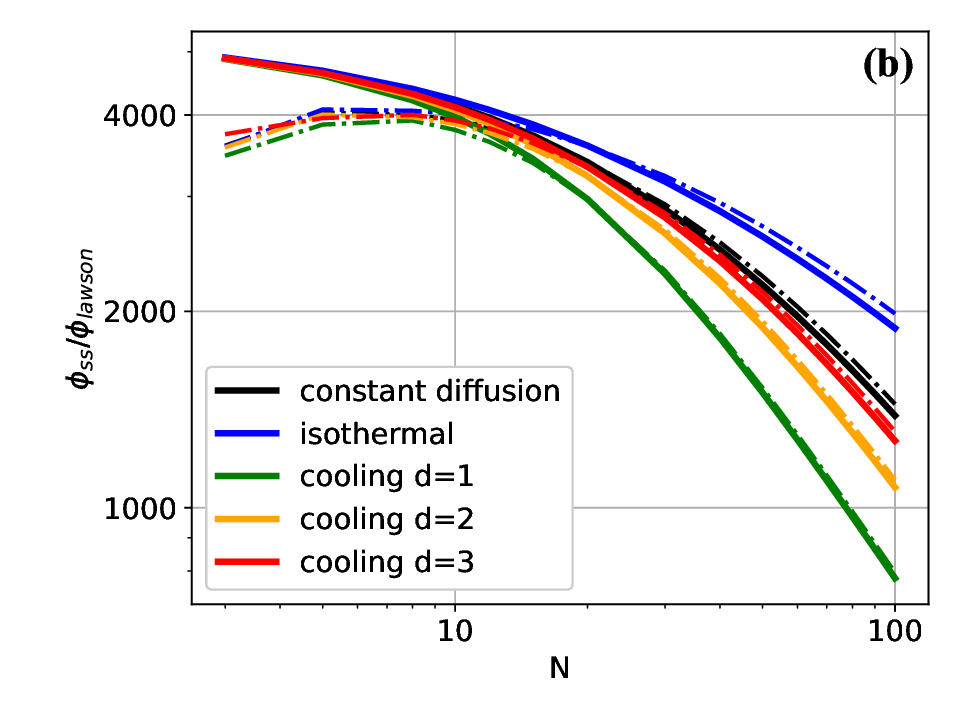}
    \caption{Steady-state flux normalized by the maximal Lawson flux $\phi_{ss}/\phi_{\rm Lawson}$ as a function of system size, for different thermodynamic scenarios (solid lines): (a) the optimal regime $\lambda/l=1$ in the first cell, (b) the sub-optimal regime $\lambda/l=20$ in the first cell. The parameters in the subplots here were as in the subplots of Fig. \ref{fig: density profiles}, respectively.
    The dash-dotted lines are the theoretical results (see Sec. \ref{theory}). }
    \label{fig: flux as a function of N}
\end{figure}

We repeated the calculation in all thermodynamic scenarios for different system sizes, $N$, in the range of  $\left(3,100\right)$, where we used the parameters of Sec. \ref{parameters}. 
In Fig. \ref{fig: flux as a function of N} we plot the steady-state flux, $\phi_{ss}$, which is the critical parameter for confinement, normalized by the Lawson flux $\phi_{\rm Lawson}$ as a function the number of MM cells $N$ for both the optimal (left) and sub-optimal (right) regimes.
As expected, the flux decreases with system size because the escaping particles have more trapping sites.
A comparison with Fig. \ref{fig: MFP profiles} reveals that the smaller the MFP, the smaller the outgoing flux, \ie  better confinement.
The worst confinement is in the isothermal scenario, and the best confinement is achieved in the $d=1$ adiabatic cooling scenario, while the diffusion scenario is in between. 
Nevertheless, the maximum difference in $\phi_{ss}$ between different thermodynamical scenarios is about one order of magnitude.
This result provides the confinement improvement scale one may expect from the cooling effect in MM systems.
Notably, despite the considerably improved confinement achieved by adding the MM section even in the adiabatic cooling scenario, the outgoing flux does not meet the Lawson criterion.
Therefore, further technological progress must be made, \eg by applying external RF fields before MM systems could be realized as a fusion reactor.

Next, let us discuss the definition of confinement (or containment) time. 
A common definition is the time required for a particle in the central cell to travel outside the system. \cite{logan1972multiple}
This definition is approximately equivalent to the total number of particles in the whole system (including those in the MM sections) divided by the (steady-state) outgoing flux.
The confinement time under this definition was found to increase quadratically with $N$ for the MM system. \cite{logan1972multiple, makhijani1974plasma} 
However, only the plasma in the central cell contributes to the fusion power since the plasma expands in the MM sections, and fusion quickly becomes negligible.
Therefore, a more relevant definition may be the time a particle spent in the central cell, which approximately equals the ratio between the number of particles in the central cell and the steady-state outgoing flux, $\phi_{ss}$.
In other words, it is the average time required for a particle in the central cell to diffuse out to the first MM cell rather than out of the whole system.
Therefore, the confinement time in this definition is expected to scale as $1/\phi_{ss}$, \ie to increase linearly with $N$.
Indeed, Fig. \ref{fig: flux as a function of N} shows that in the optimal regime (left panel), the flux decays approximately as $1/N$ in the constant diffusion and cooling scenarios, while in the isothermal scenario, the decay rate is more modest because the MFP increases as the plasma expands, making the cells less effective for particles trapping.
The sub-optimal regime (right panel) exhibits a similar picture but approaches the $1/N$ scaling only for sufficiently large $N$, demonstrating the importance of the plasma MM being in optimal regime to exploit the MM effect efficiently.

Finally, we briefly compare our rate equations model results to those of Skovordin and Beklemishev's kinetic model\cite{skovorodin2012plasma} (Fig. 6), in which the parameters were similar to those of our sub-optimal regime in the isothermal scenario. 
In both cases, we consider $R_m=10$ and $l=20\lambda$ and estimate (from the graph) the improvement when increasing $N$ from 10 to 100.
We found that the two models reasonably agree for this set of parameters, where the confinement time increases by about a factor of $2$ in both models.

\subsection{Scaling With Mirror Ratio}
\label{scaling with Rm}

\begin{figure}[tb]
    \centering
    \includegraphics[clip, width=1.0\linewidth]{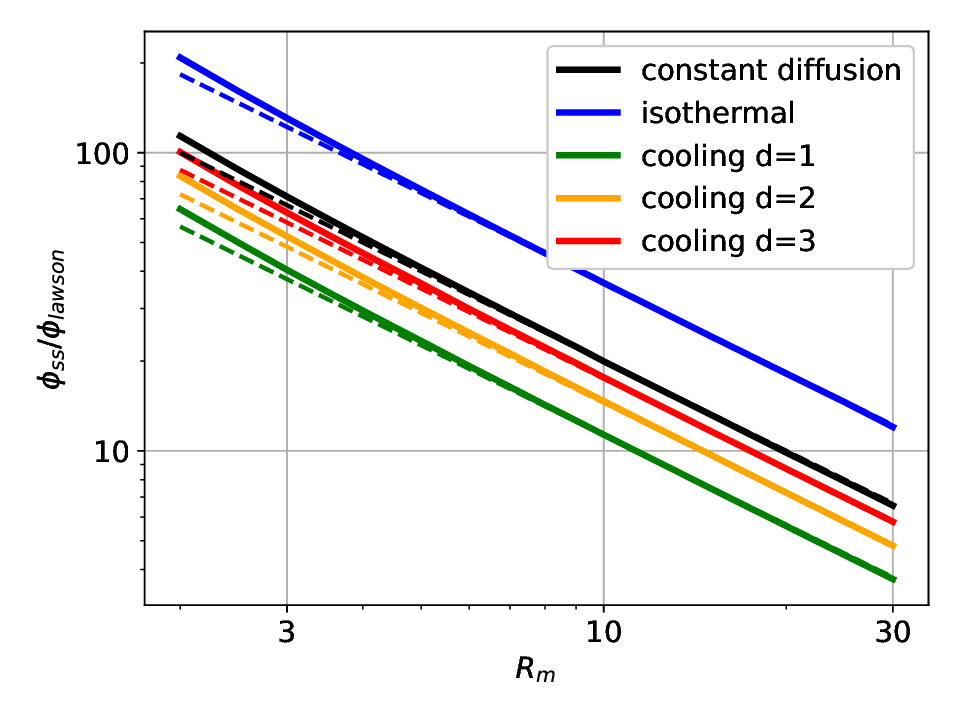}
    \caption{The steady-state fluxes, $\phi_{ss}$, normalized by the maximal Lawson flux, $\phi_{\rm Lawson}$, as a function of the mirror ratio, $R_m$, for different thermodynamic scenarios (solid lines) in the optimal regime ($\lambda/l=1$ in the first cell) are compared with a reference scaling low of $1/R_m$ (dashed lines). 
    The system parameters here were $l=10\text{m}$, $N=30$, $k_{\rm B}T=3 \text{keV}$, $B=10\text{T}$, and $n=2\cdot 10^{22}\text{m}^{-3}$.}
    \label{fig: flux as a function of Rm}
\end{figure}

The last effect we study is the scaling of $\phi_{ss}$ with the mirror ratio, $R_m$. 
We consider the optimal regime parameters (see Sec. \ref{parameters}) and $N=30$ but change the mirror ratio, $R_m$, in the range $\left(2,30\right)$. 
In Fig. \ref{fig: flux as a function of Rm}, we plot the steady-state flux as a function of the mirror ratio. 
We compare our simulation results (solid lines) with the predicted reference scaling of $1/R_m$ (dashed lines) for the (medium and long) strong-ripples regime, $(1<\lambda/l<R_m)$, as defined by Kotelnikov.\cite{kotelnikov2007new} 
Interestingly, the excellent agreement between the simulation and theory for $R_m>10$ is not only for the usually considered isothermal plasmas but also for all the other thermodynamic scenarios.
We note that this scaling also holds in the sub-optimal regime (not shown in the figure).
The deviation observed for smaller mirror ratios, $R_m<3$ can be understood by noticing that the system approaches the mild (and weak) ripple regimes in which the dependency of the confinement time ($\sim \phi_{ss}^{-1}$) on $R_m$ becomes weaker\cite{kotelnikov2007new}. 
However, a detailed study of this effect in other regimes of the parameters space is beyond the scope of the current study.

\section{Theory}
\label{theory}

As suggested by Budker et al.\cite{budker1971influence} and Danilov et al.,\cite{danilov1975plasma} when $\lambda\gg l$ (\ie the sub-optimal regime) the discrete MM system can be approximately described by a continuous, one-dimensional diffusion equation for the total density
\begin{eqnarray} \label{diffusion_eqn}
    \frac{\partial n}{\partial t} = \frac{\partial}{\partial x} \left(D\frac{\partial n}{\partial x}\right).
\end{eqnarray}
Here, $D$, is the diffusion coefficient that is generally a nonlinear function of both density and temperature, $D=D(n,T)$.
The steady-state density profile, $n(x)$, can be calculated from Eq. (\ref{diffusion_eqn}) by  solving
\begin{eqnarray}
    \frac{\partial n}{\partial t} = 0.
\end{eqnarray}
Since the typical transition time between cells is
$l/v_{th}$, we assume
\begin{eqnarray} \label{Eq: difus. coef.}
D(n,T)\approx\lambda\, v_{th}. 
\end{eqnarray}
The prefactor depends on the system's parameters, but this does not affect the steady-state density profile.
To close our model, we must add a relation between temperature and density (an equation of state).
As before, we consider three thermodynamical scenarios: (a) constant diffusion coefficient, (b) isothermal plasma, and (c) adiabatic cooling.
In addition, we consider boundary conditions for the density profile, $n(x=0)=n_{0}$ and $n(x=L)=n_{1}$.
Under these assumptions, one can derive the steady-state flux for each thermodynamical scenario via Fick's law,
\begin{eqnarray}
    \phi_{ss}=-D\frac{\partial n}{\partial x}.
\end{eqnarray}
Notably, while one of the boundary conditions, say $n_0$,  determines only the dimensions of the solution and does not affect the shape of the profile nor the steady-state flux out of the system, the second boundary condition, say $n_1$, or more accurately the dimensionless parameter, $n_1/n_0$, does.
In other words, unlike the semi-kinetic rate equations model, the analytical diffusion model predicts the density profiles but not the steady-state flux that also depends on the imposed boundary condition.
Therefore, to quantitatively compare the rate equations model and the analytical diffusion theory, we take the value of $n_1/n_0$ received from the rate equations solution for each case and substitute it in the relevant following analytical expressions.
Finally, we note that although our theory is based on the simple diffusion equation rather than more comprehensive theories for open systems,\cite{mirnov1996multiple,kotelnikov2007new} it is, as will be shown next, sufficient to describe the main effect found here, \ie the adiabatic cooling effect.

\subsection{Constant Diffusion Scenario}
Though not so physical, the simplest scenario is when both $T$ and $\lambda$ are constant and therefore the diffusion coefficient as well, $D=D_{0}$.
In this case, Eq. (\ref{diffusion_eqn}) becomes
\begin{eqnarray}
    \frac{\partial^{2}n}{\partial x^{2}}=0.
\end{eqnarray}
The solution is simply a linear density profile
\begin{eqnarray} \label{theor profile constD}
    n\left(x\right)=n_{0}\left(1-\frac{n_0-n_1}{n_0}\frac{x}{L}\right)
\end{eqnarray}
and the outgoing flux reads
\begin{eqnarray} \label{theor flux D}
    \phi_{ss}=D_{0}\frac{n_{0}-n_{1}}{L}.
\end{eqnarray}

\subsection{Isothermal Scenario}
If, as commonly considered, the MM system is isothermal, the diffusion coefficient depends only on the cell density. 
By considering constant temperature in Eq. (\ref{Eq: difus. coef.}), we can write
\begin{eqnarray}
D=D_{0}\frac{n_{0}}{n},
\end{eqnarray}
where, $D_0$ is constant.
In this scenario, the diffusion equations becomes nonlinear 
\begin{eqnarray}
    \frac{\partial}{\partial x}\left(\frac{1}{n}\frac{\partial n}{\partial x}\right)=0.
\end{eqnarray}
The solution (subjected to the same boundary condition as before) decays exponentially with $x$,
\begin{eqnarray}\label{theor profile iso}
    n\left(x\right)=n_{0}e^{-\frac{x}{L}\ln\frac{n_{0}}{n_{1}}},
\end{eqnarray}
and the outgoing flux reads 
\begin{eqnarray} \label{theor flux iso}
    \phi_{ss}=D_{0}\frac{n_{0}}{L}\ln\frac{n_{0}}{n_{1}}.
\end{eqnarray}

\subsection{Cooling Scenarios}
The last but most interesting case is the adiabatic cooling scenario.
In this regime, the temperature scales as $T\propto n^{\gamma-1}=n^{2/d}$ so the diffusion coefficient can be written as 
\begin{eqnarray}
D=D_{0}\left(\frac{n}{n_{0}}\right)^{\frac{5}{d}-1}
\end{eqnarray}
The associated non-linear diffusion equation is then
\begin{eqnarray}
    \frac{\partial}{\partial x}\left(n^{\frac{5}{d}-1}\frac{\partial n}{\partial x}\right)=0,
\end{eqnarray}
resulting in a power law steady-state profile,
\begin{eqnarray}\label{theor profile cooling}
    n\left(x\right)=n_{0}\left(1+\left[\left(\frac{n_{1}}{n_{0}}\right)^{\frac{5}{d}}-1\right]\frac{x}{L}\right)^{\frac{d}{5}},
\end{eqnarray}
where the boundary conditions are as in the previous cases.
The outgoing flux in this case is
\begin{eqnarray} \label{theor flux cooling}
    \phi_{ss}=-D_{0}\frac{n_{0}d}{5L}\left[\left(\frac{n_{1}}{n_{0}}\right)^{\frac{5}{d}}-1\right].
\end{eqnarray}
It is noted that in the limit of $d\rightarrow \infty$, the adiabatic expansion scenario approaches the isothermal scenario with an exponential steady-state density profile.

\subsection{Discussion}
In Figs \ref{fig: density profiles} and \ref{fig: flux as a function of N}, we compared our theoretical predictions with the simulation results of the rate equations model for the steady-state density profiles and the outgoing fluxes, respectively.
The comparisons include all three thermodynamical scenarios, where the theoretical profiles are developed in Eqs. (\ref{theor profile constD}),  (\ref{theor profile iso}), and (\ref{theor profile cooling}) and  theoretical fluxes (as a function of $N$) are given in Eqs. (\ref{theor flux D}), (\ref{theor flux iso}), and (\ref{theor flux cooling}).
Solid lines in the figures plot the simulation results, while dashed-dotted lines illustrate the theory. 
Our theory exhibits an excellent agreement with the simulation results in the sub-optimal regime, $\lambda\gg l$ (Fig. \ref{fig: density profiles}b), which is quite expected because one of the main assumptions in the base of the diffusion theory for MM systems is nothing but $\lambda\gg l$.
Remarkably, also in the optimal regime, where $\lambda \approx l$ (Fig. \ref{fig: density profiles}a), the theoretical density profiles agree pretty well with simulations in the isothermal, linear diffusion, and $d=3$ cooling regimes while the theory begins to deviate from the simulations for $d=1,2$.
This deviation is partially understood by looking at the MFP for the different scenarios depicted in Fig. \ref{fig: MFP profiles} noting that the MFP decreases most rapidly in these two scenarios toward $\lambda/l \ll 1$.
In contrast, in the other thermodynamical scenarios, $\lambda$ is of the order of $l$ or larger.

In addition to the outer boundary condition, $n_1$, that is taken from the rate equations numerical results for each value of $N$, the theoretical outgoing fluxes (Fig. \ref{fig: flux as a function of N}) depend on the unknown dimensional parameter, $D_0$, in the expressions for the diffusion coefficient. 
Thus, to compare the theoretical fluxes with simulations, we normalize the values of the theoretical fluxes, $\phi_{ss}$, such that they will be equal to the numerical fluxes at $N=20$.
Under this normalization, the excellent agreement between the theoretical diffusion model (dashed-dotted lines) and the numerical rate equations results (solid lines of the same color) in both optimal and sub-optimal regimes demonstrates the consistency of the theory in its validation region, $N\ge10$.

\section{Conclusions}
\label{conclusions}

In this work, we have developed a rate equations model for the MM system and solved it for the steady-state in three thermodynamic scenarios: constant diffusion, isothermal, and adiabatic expansion with $d=1,2,3$ degrees of freedom. 
Two regimes have been studied: (a) the so-called optimal regime, where $\lambda/l=1$ at the first MM cell and (b) the sub-optimal regime, where $\lambda/l=20$ at the first MM cell.
We compared the resulted steady-state axial density profiles with an analytical theory based on the diffusion equation in all three thermodynamical scenarios.
The comparison shows an excellent agreement for $\lambda/l\gg1$ (the sub-optimal regime), consistent with the theory's assumptions.
The optimal regime, $\lambda/l\approx 1$, also exhibits a good agreement for scenarios where the MFP does not decrease too rapidly. 

The main result of our model is that the thermodynamic scenario significantly affects the steady-state density profiles and the outgoing fluxes, where adiabatic cooling is the best scenario with about 5-fold longer containment time than the isothermal scenario.
It was also found that the steady-state flux, $\phi_{ss}$, inversely decays with the number of MM cells, $N$, for $N>10$  in the optimal regime and with the mirror ratio, $R_m$.\cite{kotelnikov2007new}
Remarkably, these scaling hold for all considered thermodynamic scenarios.
It is noted that our model predicts a convex declining density profile for the isothermal scenario and a concave shape for the cooling scenario.
Although it is beyond the scope of the current study to determine the thermodynamical scenario for a given MM system, our prediction regarding the density profile curvature in the different thermodynamical scenarios can, in principle, be tested experimentally. 

As a concluding remark, we recall that adding MM sections improves the confinement time significantly, especially in the adiabatic cooling regime. 
Nonetheless, the improvement resulting from the cooling effect alone is not enough to realize a fusion reactor and should be combined with other advanced control methods that exploit various kinetic and collective effects.
In this context, we point out two exciting directions, moving magnetic mirrors\cite{be2018plasma} and external RF fields.

\section*{Data Availability Statement}
The data that support the findings of this study are available from the corresponding author upon reasonable request.

\section*{Acknowledgments}
This work was supported by the PAZI Foundation, Grant No. 2020-191.


\bibliographystyle{aipnum4-1}
\bibliography{references}

\end{document}